\documentclass[conference,letterpaper]{IEEEtran}
\ifCLASSINFOpdf
\else
\fi
\usepackage[cmex10]{amsmath}
\usepackage{graphicx}

\newcommand\remove[1]{}

\begin{document}
\title{A Quantum Information Retrieval Approach to Memory}

\author{
\IEEEauthorblockN{Kirsty Kitto and Peter Bruza}
\IEEEauthorblockA{School of Information Systems\\
Queensland University of Technology, Brisbane, Australia\\
Email: [kirsty.kitto,p.bruza]@qut.edu.au}
\and
\IEEEauthorblockN{Liane Gabora}
\IEEEauthorblockA{Psychology and Computer Science\\
University of British Columbia, Kelowna, Canada \\
Email: liane.gabora@ubc.ca}}

\maketitle

\begin{abstract}
%\boldmath
As computers approach the physical limits of information storable in memory, new methods will be needed to further improve information storage and retrieval. We propose a quantum inspired vector based approach, which offers a contextually dependent mapping from the subsymbolic to the symbolic representations of information. If implemented computationally, this approach would provide exceptionally high density of information storage, without the traditionally required physical increase in storage capacity. The approach is inspired by the structure of human memory and incorporates elements of G\"{a}rdenfors' Conceptual Space approach and Humphreys et al.'s matrix model of memory. 
\end{abstract}

\IEEEpeerreviewmaketitle

\section{Memory Structure and Information Density}

The age of density driven computer memory increase is fast approaching its conclusion. With Moore's law suggesting that we are nearing the physical limits of information density storable in standard computational memory, it is time to investigate new paradigms of information storage and retrieval. This paper proposes that a recently developed class of cognitive models provide a highly promising avenue, that can be used to shift the current information storage paradigm from a density dependent model to a more structural methodology. 
Our approach is inspired by insights from neuroscientific studies of memory, and focuses upon the context in which  information is encoded and subsequently recalled. Mathematically, it is grounded in a vector-based formalism that utilizes the probability structure of quantum theory which draws upon two related lines of research. One derives from modern approaches to information retrieval which attempt to incorporate a sophisticated notion of context into the classification of information as relevant to a query (\cite{vanRijsbergen:geometry,widdows:geometry, song.bruza:towards, bruza.widdows.ea:quantum}). The second approach is more squarely based in cognitive science, and uses a quantum approach to model concepts and their combinations (\cite{aerts.gabora:theory, aerts.gabora:theoryII, kitto.ramm.ea:testing,kitto.ramm.ea:quantum}).
% LMG This para was moved up to intro in response to critique from Reviewer 4

In summary, the key purpose of this paper is to suggest a new paradigm for information storage and retrieval \emph{in context} that allows for a marked increase in the amount of information storable by a given resource. This will require the identification of a mechanism by which stored information can be retrieved, which somehow links that information to relevant storage and retrieval contexts. We provide tentative solutions for both of these problems.  

We begin with a brief summary of how a subsymbolic encoding in human memory can still give rise to a symbolic capacity. This is followed by a review of the Conceptual Space approach advocated by G\"{a}rdenfors \cite{gardenfors:conceptual}, which proposes a framework of three tiers for understanding human memory. We then discuss the Matrix Model of Memory \cite{humphreys.bain.ea:different}, which shows how a memory can be encoded along with information about the context in which it occurred. This will lead us to consider the treatment of context in that model and finally to extend it through reference to a quantum information retrieval framework which combines the desirable features of each approach. We propose that our approach not only allows for an exceptionally high density memory storage but also provides a memory architecture that can process information in a way that is flexible, adaptive, and possibly even creative. 

\section{Symbolic and Subsymbolic Levels of Human Memory}

Let us begin by examining the architecture of human memory (summarized in \cite{gabora:revenge}). This will serve as a starting point to build a computer memory that uses similar basic mechanisms to human memory. 

\subsection{The Subsymbolic Level}
% This section was modified by LMG in response to critique by reviewer #3
We take as a starting point some fairly well established characteristics of memory. Human memories are encoded in neurons that are sensitive to ranges (or values) of what has been called subsymbolic microfeatures \cite{smolensky:on,churchland.sejnowski:computational}. For example, one might respond to lines of a particular orientation, or the quality of honesty, or quite possibly something that does not exactly match an established term \cite{mikkulainen:neural}. Note that sometimes use the word concept is used by non-neuroscientists (\emph{e.g.} \cite{roy:discovery}) to refer to subsymbolic microfeatures. In this paper, the word microfeatures is used to refer to stimuli responded to by single cells, which may or may not be meaningful in daily life, and the word concepts to refer to things like DOG or BEAUTY that are generally comprised of many microfeatures, and refer collectively to a class of instances or exemplars that are meaningful in daily life. 

Another characteristic of memory is that although each neuron responds maximally to a particular microfeature, it responds to a lesser extent to related microfeatures, an organizational structure referred to as coarse coding \cite{hubel.wiesel:receptive}. For example, neuron A may respond preferentially to sounds of a certain frequency, while its neighbor B responds preferentially to sounds of a slightly different frequency, and so forth. However, although A responds maximally to sounds of one frequency, it responds to a lesser degree to sounds of a similar frequency. The upshot is that an item in memory is stored in a distributed manner across a cell assembly that contains many neurons, and likewise, each neuron participates in the storage of many items \cite{hebb:organization}. A given experience activates not just one neuron, nor every neuron to an equal degree, but activation is spread across members of an assembly. This means that the same neurons get used and re-used in different capacities, a phenomenon referred to as neural re-entrance \cite{edelman:neural}. 

The final key attribute of memory is its content addressability, meaning that there is a systematic relationship between the content of a representation, and the neurons where it gets encoded. This emerges naturally as a consequence of the fact that representations activate neurons that are tuned to respond to particular features, so representations that get encoded in overlapping regions of memory share features. As a result, they can thereafter be evoked by stimuli that are similar or \emph{resonant} in some (perhaps context-specific) way \cite{hebb:organization,marr:theory}. Note that even if a brain does not possess a neuron that is maximally tuned to a particular microfeature, the brain is still able to encode stimuli in which that microfeature predominates, because representations are distributed across many neurons. 

Note that on the basis of the discovery of single cells in the human brain that have highly selective, abstract and invariant responses to complex, natural stimuli, which have unfortunately been called concept cells, some neuroscientists have questioned the idea that representations are distributed \cite{roy:discovery}. This is not inconsistent with the variety of distributed representation discussed here. If you artificially activate one neuron, it gives an invariant response. It is because real-world stimuli and experiences activate not just one neuron but many that actual representations in memory are distributed.

\subsection{The Symbolic Level}

Consciously experienced symbolic meanings emerge in response to the set of subsymbolic microfeatures responded to by the entire constellation of activated neurons. Sometimes these neurons have been activated as a unit many times before, at others the constellation consists of neurons that have never been activated simultaneously as a whole. In this case an emergent meaning may simultaneously incorporate elements of different symbolic representations. 

The distributed, content addressable architecture of memory is critical to the adaptive, flexible, and creative manner in which the information it stores is not just retrieved when required, but frequently reconstructed in contextually appropriate and sometimes even creative ways. If this memory were not distributed then there would be no overlap between items that share microfeatures, and thus no means of forging associations between them. If it were not content-addressable then these associations would not be meaningful. The upshot is that representations which share features are encoded in overlapping distributions of neurons, and therefore activation can spread from one to another. Thus representations are encoded in memory in a way that takes into account how they are related, even if this relationship has never been consciously noticed \cite{gabora:cognitive,gabora:revenge,gabora.ranjan:how}.
This is not earth shattering; indeed it seems fairly obvious with respect to the hierarchical structure of knowledge. We may never have explicitly learned that a white hamster is a mammal, but we know it is one nonetheless. In this sense it is reasonable to claim that people implicitly know more than they have ever explicitly learnt. This architecture has implications that extend far beyond issues related to the hierarchical structure of knowledge. 

It should be pointed out how different this is from the typical structure of computer memory. In a computer memory, each possible input is stored in a unique address. Retrieval is thus a matter of looking up the address in a register and then fetching the corresponding item at the specified location. Since there is no overlap of representations, there is no means of creatively forging new associations based on newly perceived similarities. The exceptions are computer architectures that are designed to mimic, or are inspired by, the distributed, content-addressable nature of human memory, but these are difficult to discuss formally. In this paper we shall propose a theoretical structure that can be used to map subsymbolic architectures to symbolic representations, and so potentially provide a more flexible, adaptable and creative approach to computer memory.

\subsection{Forging Unusual Associations through Reconstructive Interference of Memories}

A fascinating finding to come out of the early connectionist literature is that in a distributed, content addressable memory, not only do representations that share features activate each other, they sometimes interact in a way that is creative. Even a simple neural network is able to abstract a prototype, fill in missing features of a noisy or incomplete pattern, or create a new pattern on the fly that is more appropriate to the situation than anything it has ever been fed as input \cite{mcclelland.rumelhart:distributed}. In fact, similar representations can interfere with one another \cite{feldman.ballard:connectionist,hopfield:neural,hopfield.feinstein.ea:unlearning}, and these same papers provide numerous names for this phenomenon: crosstalk; false memories; spurious memories; ghosts; and superposition catastrophe.
These phenomena are suggestive of a form of thought that, if not outright creative, involves a departure from known reality. Findings from neuroscience are also highly consistent with this phenomenon; as Edelman puts it, one does not retrieve a stored item from memory so much as reconstruct it \cite{edelman:bright}. That is, an item in memory is never re-experienced in exactly the form it was first experienced, but colored, however subtly, by what has been experienced in the meantime, re-assembled spontaneously in a way that relates to the task at hand (one reason eye-witness accounts cannot always be trusted \cite{paterson.kemp.ea:co-witnesses,loftus:memory,schacter:seven}).

Because information is encoded in a distributed manner across ensembles of neurons interacting by way of synapses, the meaning of a representation is in part derived from the meanings of other representations that excite similar constellations of neurons; that is why memory is sometimes referred to as associative. Content addressability ensures that the brain naturally brings to mind items that are similar in some perhaps unexpected or indefinable but useful or appealing way to what is currently being experienced. Recall that if the regions in memory where two distributed representations are encoded overlap then they share one or more microfeatures. They may have been encoded at different times, under different circumstances, and the correlation between them never explicitly noticed. But the fact that their distributions overlap means that some context could come along for which this overlap would be relevant, causing one to evoke the other. There are as many routes by which an association between two representations can be forged as there are microfeatures by which they overlap; i.e., there is room for typical as well as atypical connections. Therefore what gets evoked in a given situation is relevant, and that happens for free; no search is necessary at all because memory is content-addressable. The \emph{like attracts like} principle is embedded in our neural architecture. 

Moreover, because memory is distributed and subject to crosstalk, if a situation does come along that is relevant to multiple representations, they merge together, a phenomenon that has been termed  \emph{reconstructive interference} \cite{gabora.saab:creative}. The multiple items may be so similar to each other that you never detect that the recollection is actually a blend of many items, and in this case the distributions of neurons activated overlap substantially. Alternatively, they may differ in mundane ways, as in everyday mind-wandering. They may be superficially different but related in a way you never noticed before, in which case the distributions of neurons activated overlaps only with respect to only a few features that happen to be relevant or important in the present context. Finally, the present experience may infuse recall of a previous experience that is relevant or important with respect to only a few key features.

We now turn to some of the models that have been proposed to describe this merging of the subsymbolic and the symbolic levels of human memory. We shall find a number which share a key set of features.

\section{Memory Models Inspired by the Multi-leveled Architecture of Human Memory}

G\"{a}rdenfors \cite{gardenfors:conceptual} has proposed a three level model of cognition in which the representation of information varies greatly at each level. Within the lowest level, information is pre-conceptual or subsymbolic, and is carried by a connectionist representation.  

At the uppermost level information is represented in terms of higher order symbolic structures such as sentences. Grammars specify the parts of a sentence, and the manner in which they fit together. It is at this upper  symbolic level of cognition where  a significant portion of the computational literature resides. Indeed, the very storage of information in a standard computer architecture could be understood as belonging to this level. 

While the need for at least these two levels seems intuitively plausible, the gap between the upper, logical level and the lowest connectionist level is difficult to bridge. How are we to connect the symbolic approaches with the structural? There is a possibility for some approach that shares logical \emph{and} structural characteristics \emph{between} the symbolic and the structural levels of cognition, and this is precisely where G\"{a}rdenfors' intermediate, conceptual level, or \emph{conceptual space},  is introduced. Rather than relying upon a connectionist structure, an intermediate geometric representation is used which provides an expressive theoretical framework capable of linking the `hardware' of a `neuronal' level with the more commonly described, and theoretically understood, logical level.

\section{Encoding Information in a Conceptual Structure}
\label{sec:geometric}

Within a conceptual space, knowledge has a dimensional structure. For
example, the property COLOR can be represented in terms of three dimensions: hue, chromaticity, and brightness, which can be mapped into a convex region in a geometric space. Thus, the property RED is a convex region within the tri-dimensional
space made up of hue, chromaticity and brightness, and the property BLUE
would occupy a different region of this same space. A domain is a set of integral dimensions in the sense that values in particular dimensions can determine (or
affect) the values possible in others. For example, the three dimensions defining the color space are integral since the brightness of a color
will affect both its saturation (chromaticity) and hue. G\"ardenfors extends
the notion of properties into concepts, which are likewise based on domains. 

For example, the concept APPLE may have domains taste, shape, color, etc. Context is
modeled as a weighting function on the domains, for example, when eating
an apple, the taste domain will be prominent, but when playing with it,
the shape domain  (i.e. its roundness) will be heavily weighted. 

Observe the distinction between representations at the symbolic and conceptual levels. At the symbolic level, the concept APPLE can be represented as the
atomic proposition apple(x). However, within a conceptual space (conceptual level), it has a representation involving multiple inter-related dimensions and domains. Colloquially speaking, the  token ``apple'' (which might be spoken, written \emph{etc.}) is the tip of an iceberg with a rich underlying representation at the conceptual level. G\"ardenfors points out that the symbolic and conceptual
representations of information are not in conflict with each other, but are
to be seen as ``different perspectives on how information is described''.

However, an implementation problem arises in that both the representation and the generation of a conceptual space from its underlying content has generally been discussed only for simple examples such as those above. It is not clear how more complex examples could be implemented. A more comprehensive and systematic approach to the representation of conceptual spaces is required.

Vector space based models (VSBM) provide a viable first avenue here. These can be traced back to the seminal paper of Salton
et al. \cite{salton.wong.ea:vector} who were searching for an appropriate
mathematical space to represent documents for the purposes of Information Retrieval. Starting from a few basic
desiderata, they settled upon a vector in a high dimensional vector space as
an appropriate representation of a document. Within this framework, a query
is treated like a small (pseudo) document that is also converted to vector
form. The documents in the corpus are then ranked according to their
distance to the query; closer documents are considered more relevant than
ones that are further away. One of the main drawbacks of this system was that it had trouble
returning documents that would have been highly relevant if one of the words
in the query was replaced by a synonym. The next advance came from
representing concepts latently in a so-called \emph{semantic space} where
they are not formally represented or labeled. Semantic spaces are instances
of vector spaces, and represent words in a basis created from other words,
concepts, documents, or topics. They are generally built from the
observation of co-occurrences in large text corpora. In word spaces such as
the Hyperspace Analogue to Language (HAL) \cite{schutze:automatic}, the basis consists of
every word in the vocabulary. Thus, the vector for a given word $w$ is
calculated by summing the number of occurrences of word $w(i)$ in a given
context window around each occurrence of $w$ and writing that number at the
position $i$ in the vector that represents $w$. This number can be adjusted
using the distance (defined in terms of the number of words) or mutual
information measures such as Point-Wise Mutual Information, which allows for
a weighting of the importance of the word at that position. It is also
possible to take word order into account \cite{jones.mewhort:representing,sahlgren.holst.ea:permutations}. Later models derived  a more fundamental semantic value through a reduction of the initial
word space using mathematical tools such as Singular Value Decomposition 
\cite{landauer.dumais:solution}, Non Negative Matrix factorization \cite{lee.seung:learning}, 
or random projection \cite{sahlgren:introduction}, all of which generate a new smaller basis which can under certain conditions be naturally related to certain topics, objects and
concepts \cite{lee.seung:learning}. 

Semantic space models, however, do not make provision for integral dimensions (a notion related to that of `core properties' of a concept). 
This leaves them too situation dependent, and relevant primarily for the text collection from which they were constructed. For the purposes of next generation information storage, we will require a more objective information storage mechanism that can function satisfactorily at the conceptual level. Thus, while learning from the semantic space approaches, this paper will propose that we extend them from a text based, and corpus dependent information representation, to a concept and property inspired approach. However, the vector based analytical contributions of the semantic space approaches will play a key inspirational role as we start with this extension.

\section{ Information Retrieval in a Matrix Model of Memory}
\label{sec:mm}

We now turn to a promising treatment of retrieval, that is capable of providing a map between the neural and conceptual levels of information storage. The matrix model of memory \cite{humphreys.bain.ea:different} is a well known cognitive model. It stores and encodes memories as patterns of interconnections between the elements that define items in memory (\emph{i.e.} the  neurons for a subsymbolic structure). All memories are superimposed (summated) in this representation so that, without appropriate
cuing, their individual identities are lost. Thus, the model provides a natural link between the lower and mid levels of information that G\"ardenfors proposes. For example, when a set of interconnected neurons fires, this can be represented in the matrix model as a set of entries in a matrix, with the entries in the matrix corresponding to the probability that a particular pairing of nodes will concurrently fire (although this is not a necessary interpretation of the model \cite{humphreys.bain.ea:different}).

Humphreys et al. \cite{humphreys.bain.ea:different} take the position that there are two fundamental but pervasive memory access operations; matching and retrieval. 
Matching involves the comparison of test cue(s) with the information stored in memory, and gives the strength of 
the match as output. In contrast, retrieval involves the recovery of an associate of a cue (i.e. the return of actual information), and so is the concept in which we are currently  interested.

The matrix model takes an item $A_i$, occurring in a context $X$, to retrieve a list associate $B_i$. This assumes that a three-way association (between the context, the cue and the desired item) must have been stored. This is represented mathematically as the three-dimensional array $\bf xa^{'}_ib_i^{''}$, where $\bf x$ is a column vector, $\bf a_i^{'}$ is a row vector, $\bf xa_i^{'}$ is a $n\times n$ matrix, and $\bf b_i^{''}$ is another vector in an orthogonal dimension to $\bf x$ and $\bf a_i^{'}$. (Primes are used to indicate this set of orthogonality relationships.)
The matrix $\bf xa_i^{'}$ represents the \emph{association} between the context $\bf x$ and the cue $\bf a_i^{'}$.
If a list of items $A_1B_1, A_2B_2 \dots A_kB_k$ is learned in a context $X$, then Humphreys et al. define the memory for the list as a simple sum over all the three-dimensional arrays that were formed:
\begin{equation}
 {\bf E}=\sum_{i=1}^k {\bf xa^{'}_ib_i^{''}}.
\end{equation}
This list memory $\bf E$ is added onto any pre-existing memories in a process that we leave to the original article \cite{humphreys.bain.ea:different}. 

Retrieval from this list memory is defined by Humphreys et al. \cite{humphreys.bain.ea:different} to work as follows. First, a test cue $\bf x a_j^{'}$ is applied to the list memory:
\begin{align}
( {\bf x a_j^{'}})\cdot {\bf E} &= ({\bf x a_j^{'}})\cdot\left(\sum_{i=1}^k {\bf xa^{'}_ib_i^{''}}\right)\\
&= \sum_{i=1}^k [{\bf ( x a_j^{'})\cdot (xa^{'}_i})]{\bf b_i^{''}}\\
&= \sum_{i=1}^k [{\bf ( x \cdot x ) (a_j^{'}\cdot a^{'}_i})]{\bf b_i^{''}}\\
&=  [{\bf ( x \cdot x ) (a_j^{'}\cdot a^{'}_j})]{\bf b_j^{''}}+ \sum_{i\neq j} [{\bf ( x \cdot x ) (a_j^{'}\cdot a^{'}_i})]{\bf b_i^{''}}.\label{eq:retrieval}
\end{align}
%he result of this retrieval process is a vector that can be represented as a weighted composite of the correct target vector, $\bf b_j^{''}$, and the other target vectors in the list ${\bf b_i^{''}}, (i\neq j)$. 
We can learn a little about this model through a consideration of the two terms in \eqref{eq:retrieval}. The first term represents the desired vector $\bf b_j^{''}$, weighted by a scalar term that results from taking the dot products of two vectors with themselves. The second term is effectively an error term; if the similarity between the cue $\bf a_j^{'}$ and the other cues that were used to store the memory ($\bf a_i^{'}$) is too great then this error term will become large and the chances of the correct item being recalled will decrease. In short, the other stored memories (${\bf b_i^{''}}, (i\neq j)$) will interfere with the desired term. It is also worth noting that the explicit representation of the context vectors using a dot product (i.e. $\bf x\cdot x$) in \eqref{eq:retrieval} suggests that the authors were open to the idea of a different context being present during the recall process, even if this was not explicitly discussed \cite{humphreys.bain.ea:different}. We shall return to this point in section~\ref{sec:remembering}, however, a brief foreshadowing of that argument runs as follows: 
We consider the use of context in this model to be unsatisfactory. Firstly, while the role of context is fundamental to this model, it must be explicitly and fully recorded at the time of storage. A slightly different context, or even a more detailed specification of the same context could result in the retrieval of a very different piece of information, even though a very similar cue was utilized.
The static treatment of context that is provided by this model therefore leaves us with what i an interesting retrieval paradigm, which is however perhaps unnecessarily limited. We are left wondering if there is scope for a more adaptive treatment of context, one provided by the geometric models of conceptual space that were introduced in section~\ref{sec:geometric}. 

In the remainder of this article we shall endeavor to connect these two interesting approaches into an integrated cognitive memory model which could be used to form the basis of a future physical implementation of computational memory.
This approach takes its inspiration from a set of models that consider information retrieval in context, utilizing the powerful formalism of quantum theory \cite{vanRijsbergen:geometry}--\cite{bruza.widdows.ea:quantum,bruza.cole:quantum,melucci:basis}. 

\section{Incorporating Context into Information Encoding and Retrieval}
\label{sec:qconceptual}

The seminal book by van Rijsbergen \cite{vanRijsbergen:geometry} provides a novel approach to the modeling of semantic spaces, inspired by Quantum Theory (QT). This approach models a word $w$ as a vector
\begin{equation}
\label{eq:ket}
 |w\rangle=\left( 
\begin{array}{c}
 w_1 \\
 \vdots\\
 w_n
\end{array}
\right)
\end{equation}
where $|w\rangle$ is termed a \emph{ket}, in contrast to the row vector obtained by taking the transpose: $\langle w | = |w \rangle^T = (w_1, \dots, w_n)$. In this case, we shall take the subcomponents $w_1\dots w_n$ to be the weights allocated to each of the available \emph{senses} that the word might take in a $n$-dimensional vector space. 

We can quickly see the connection to both vector space based approaches and the Matrix Model of Memory. In both cases a vector was obtained, (although in each case this was via a different process) and formed the basis of further analysis. However, the formalism of quantum theory provides an extra level of structure that implicitly incorporates a more adaptive notion of context into information recall.

This is done by seriously considering what it means to define a \emph{basis} for a conceptual space. Thus, the vector of \eqref{eq:ket} must be considered \emph{in its context}; it is a representation of information within a high dimensional vector space, with the vector entries determining the extent of the vector in each of the relevant dimensions. 

\begin{figure}[h]
\centering
 \includegraphics[width=6cm]{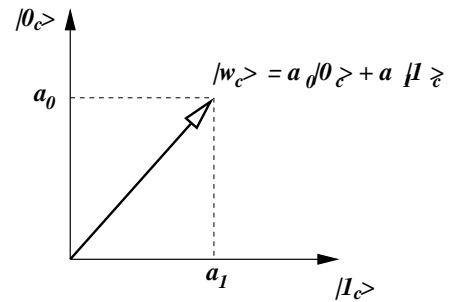}
\caption{A concept $w$, for example \emph{red}, is represented in some contextual probe $c$ which takes the form of a choice of basis. This low dimensional representation shows a case of some object being classified by a probe as ``red'' ($|1\rangle$) or ``not red''  ($|0\rangle$), with the probability of that a  ``red'' judgment being given by $|a_1|^2$. In a higher number of dimensions the property of redness will be enclosed by a convex region, and the probabilities lie in a range of values specified by the extent of that region.}\label{fig:red}
\end{figure}

This geometric model provides predictions about the likely recall of an item from memory within a given context. This is achieved via an application of Pythagoras' theorem.  Thus, simplifying equation~\eqref{eq:ket} down to a vector occurring in a two dimensional space, we might find that it could be drawn as shown in figure~\ref{fig:red}, where
\begin{align}
 \label{eq:a1wrtp}
|w\rangle =& a_0 \left(
\begin{array}{c}
0 \\
1
\end{array} \right) + a_1 \left(
\begin{array}{c}
1 \\
0
\end{array} \right)\\
=& a_0 | 0_p  \rangle+ a_1| 1_p \rangle, \mbox{ and } |a_0|^2+|a_1|^2=1.
\end{align}
Here, $\{ | 0_p  \rangle\equiv(0,1)^T, | 1_p  \rangle\equiv(1,0)^T\}$ define an orthonormal basis, and so the inner product of these basis vectors returns 0 or 1: $ \langle 0_p| 0_p  \rangle= \langle 1_p| 1_p  \rangle=1$ and $ \langle 1_p| 0_p  \rangle= \langle 0_p| 1_p  \rangle=0$. The state \eqref{eq:ket}, can be re-written using an extension of this formalism, giving
\begin{equation}
|w\rangle=w_1\left( 
\begin{array}{c}
1 \\
0 \\
 \vdots\\
0
\end{array}
\right)
+w_2
\left(
\begin{array}{c}
0 \\
1 \\
 \vdots\\
0
\end{array}
\right)
+\dots
+ w_n
\left(
\begin{array}{c}
0 \\
0 \\
 \vdots\\
1
\end{array}
\right)
\end{equation}
which allows us to capture high dimensional vector representations of information. Here, the $w_j$'s, or weights, represent the extent to which a piece of information falls into each of the dimensions of the vector space, and thus how much it overlaps with the individual concepts represented by each axis in the basis. This means that the convex region representing a property in a conceptual space can be mapped out by a collection of vectors covering that region, with each of the weights mapping how much a given property is represented by that dimension. 
A piece of stored information, (e.g. a concept $w$) is thus represented in this framework as a state, or a vector in a high dimensional space. For a low dimensional example, consider the concept of ``redness'' that might be stored about two different objects, such as an apple, and some wine. In a two dimensional, or \emph{q-bit} representation, each object will be classified as either ``red'' or ``not red'' but this classification will depend upon the context. Figure~\ref{fig:red} depicts a possible state which one of these objects might have, within a particular concept space where $|1\rangle$ represents ``red'' and $|0\rangle$ ``not red''. Within this specific context, we might find that an apple is more likely to be returned as a response to a query that asks for a ``red object'' than is red wine, although this might change were the information to be sought in a different context. We shall return to this point shortly, showing how this formalism can very naturally capture this behavior.

\begin{figure}[h]
\centering
 \includegraphics[width=6cm]{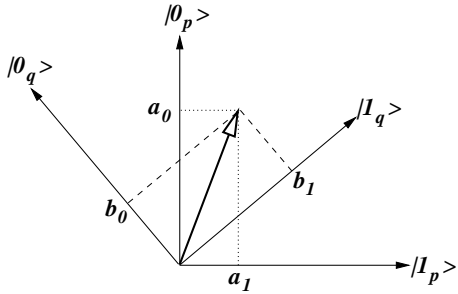}
\caption{ Changing the context of a probe can significantly affect the chances of recall. Once this effect is incorporated into a Matrix-like memory model, a structural information storage system becomes more viable.}\label{fig:2red}
\end{figure}

We propose that once information is stored in this complex multidimensional space, it can be recovered through use of a probe which enacts a \emph{quantum measurement} of the state \eqref{eq:a1wrtp}. This is defined with respect to a projection operator $V$, where, for the two dimensional case outlined above
\begin{equation}\label{eq:projector}
 V=|0_p\rangle\langle 0_p| + |1_p\rangle\langle 1_p|=V_0+V_1.
\end{equation}
According to the quantum formalism, the probability of a probe represented by the $p$ basis returning the desired value is given by
\begin{align}
 \label{eq:Aprob}
P(|1\rangle)= & \langle w | V_1 | w \rangle\\
 = & \langle w | 1_p \rangle \langle 1_p | w \rangle\\
 = & \big( a_0^*\langle 0_p | 1_p\rangle + a_1^* \langle 1_p | 1_p \rangle\big)\times\nonumber\\
  & \qquad\big( a_0 \langle 0_p | 1_p \rangle + a_1\langle 1_p | 1_p \rangle\big)\\
 = &|a_1|^2.
\end{align}
However, in the context represented by the shifted basis in figure~\ref{fig:2red} the probe returns the desired information with a probability given by $P(|1\rangle)=|b_1|^2$. 
Thus, a search for a ``red object'' in the context represented by $p$ may yield a very different result to that which searches in the context $q$.

The assumption in \eqref{eq:a1wrtp} that the squared coefficients of the basis vectors sum to 1 allows for the treatment of these values as probabilities since $0\leq P\leq1$ etc.
This approach makes use of a geometrical notion of probability \cite{isham:quantum}, which contrasts with standard probability theory, where probabilistic outcomes arise from our lack of knowledge as to what has actually occurred.
Quantum probabilities are profoundly different, arising from a genuine state of uncertainty; the context in which the information is to be represented must be defined before the recall can start to make sense.

We shall now show how this more sophisticated treatment of context can be utilized in an extension of the Matrix Memory Model which, while retaining the representation of items and cues as vectors, embeds them within a context that is spatial rather than vectorial.

\section{Remembering as a Process of Information Retrieval}
\label{sec:remembering}

In line with the proposal by Wiles et. al \cite{wiles.halford.ea:tensor}, we take the position  that the recall of information from a memory structure can be well represented by a contextual probe to an underlying network structure. The construction of such a probe has been a difficult problem for neural modelers, as it is difficult to correlate the activation of neural connections with a logical, or even conceptual, structure. However, with the three tier model advocated by G\"ardenfors we can begin to see how a probe that has both a logical (symbolic) structure, and a connection to the lower (subsymbolic) level neural model can be created. This section will sketch out the key details of this construction.

We start with a reference to the result shown in van Rijsbergen that projection operators such as the one in \eqref{eq:projector} can be used to define a conditional logic \cite{vanRijsbergen:geometry}, meaning that the link between the quantum conceptual space that we discussed in section~\ref{sec:qconceptual} and higher order logic has already been found.

This leaves the connection between the subsymbolic neural level and the conceptual levels  to be made. Returning to the consideration of the Matrix Model that was started in section~\ref{sec:mm} we recall its use of a somewhat unsatisfactory \emph{explicit context}. The representation of context in this model as a vector (i.e. $\bf x$) means that it acquires an ontological status equivalent to that of the items that are used as cues, or stored to be retrieved by those cues, however, we believe that this identification is incorrect.
%EXPLAIN WHY
Rather then behaving as a thing, or absolute entity, context appears to be more of a \emph{relationship} between the thing currently under consideration (\emph{i.e.} the memory for this scenario) and a \emph{perspective} from which it will be viewed. This is a very new approach to the treatment of context in computational representations, which most commonly take context to be a thing, or a parameter \cite{brezillon:context,guha.mccarthy:varieties} with a similar ontological status as very system which is being considered \emph{within that context}. This is unlikely to be a satisfactory approach, but the lack of alternative formal models has hindered the adoption of a more sophisticated understanding. 
However, the quantum inspired model presented above makes use of a very different conceptualization, that we shall refer to here as an \emph{implicit} context, which frames the system under consideration rather than being of a similar form to it. In what follows, we shall make use of this implicit notion of context in an extension to the Matrix Model of Memory which treats context as a space rather than a vector.

This will be achieved by utilizing projection operators rather than vectors to represent the context in which storage and recall takes place. Thus, in place of the context vector $\bf x$ in \eqref{eq:retrieval}, we propose to utilize a projection operator that arises in the same space as the memory itself
\begin{equation}
\label{eq:Vcontext}
{\bf V_x}=\sum_{h=1}^n { |x_h\rangle \langle x_h|}.
\end{equation}
This equation takes the vector notion of context utilized in equation~\eqref{eq:retrieval} and translates it into a set of projection operators defined using basis vectors, each of which could have been a context vector in the Matrix Model.
%However, this extension draws attention to an interesting feature of the Matrix Model;  the context, cue and item to be recalled are all written in different subspaces for that equation. This is why the dot products of the individual vectors in the equation only apply to variables of the same `priming type' (\emph{i.e.} none, $'$ and $''$) and are then considered separately. To us, this seems an unreasonable assumption both mathematically and psychologically. Indeed, we consider it likely that the vectors representing human memories in Humphreys et. al's Matrix Model \cite{humphreys.bain.ea:different} should be modeled in the same vector space as their cues and contexts. Memories are unlikely to be separable from those phenomena that invoke them \cite{kitto.bruza:tests,bruza.kitto.ea:non-decomposability}, indeed, the very psychological literature discussed by Humphreys et. al \cite{humphreys.bain.ea:different} suggests that the memory recalled will depend heavily upon cues and contexts. The quantum inspired model introduced in section~\ref{sec:qconceptual} offers an obvious alternative to the paradigm used by the standard Matrix Model \cite{humphreys.bain.ea:different}; all three observables (cues, contexts, and memories) occur in the same space in a well defined manner. With the definition of context as a subspace that is given above, we are now offered an option for considering cues and memories in a conceptual space which also contains the currently relevant context as specified by \eqref{eq:Vcontext}.
Returning to equation~\eqref{eq:retrieval} we rewrite it with this extended understanding of the context of a memory:
\begin{align}
{\bf V_y a_j E} &= {\bf V_y a_j }\sum_{i=1}^k {\bf V_x a_i b_i}
%\\&=\sum_{h=1}^n\sum_{i=1}^k {\bf |y_h\rangle \langle y_h| } {\bf |x_i\rangle \langle x_i| a_j^{i} a_i^{'} b_i^{''}}\\
\end{align}
where $\bf V_y$ is a second cueing context defined with respect to the vector $\bf y$ which could be specified with a different set of basis vectors to $\bf x$. Expanding the projection operators in this equation starts to give us some indication of how this  model can be expected to behave:
\begin{align}
{\bf V_y a_j E} &= \sum_{h=1}^n\sum_{i=1}^k { |y_h\rangle \langle y_h  | a_j\rangle |x_i\rangle \langle x_i| a_i\rangle | b_i\rangle}\\
&= \sum_{h=1}^n\sum_{i=1}^k { u_{hj}v_i|y_h\rangle  |x_i\rangle  | b_i\rangle} \\
&= \sum_{h,i}u_{hj}v_i {\bf y_h x_i b_i},\label{eq:Vrecall}
\end{align}
where $u_{hj}=\langle y_h  | a_j\rangle$ and $v_i=\langle x_i| a_i\rangle$ are scalars, obtained by taking the associated dot products of the corresponding vectors.  These scalars weight the contribution of the individual cue, context and stored item vectors. 
If the context of recall is the same as the context of recording (\emph{i.e.} $\bf y=x$) then we can say a little more about the recall process using a standard property of projection operators: ${\bf V_x  V_x}={\bf V_x}$ \cite{isham:quantum}.
\begin{align}
{\bf V_x a_j E} &= {\bf V_x a_j }\sum_{i=1}^k {\bf V_x a_i b_i}\\
%&={\bf V_x  V_x a_j }\sum_{i=1}^k {\bf a_i b_i}\\
&={\bf V_x   a_j }\sum_{i=1}^k {\bf a_i b_i}\\
%&=\sum_{i=1}^k { v_i|x_i\rangle|a_i\rangle|b_i\rangle}\\
&=\sum_{i=1}^k v_i{\bf x_i a_i b_i}.\label{eq:Vrecall1}
\end{align}
Finally, breaking \eqref{eq:Vrecall1} into the two components utilized in \eqref{eq:retrieval} we find that
\begin{align}
{\bf V_y a_j E} &= v_j {\bf  x_j a_j b_j}+\sum_{i\neq j}v_i {\bf  x_i a_i b_i}
\end{align}
which has the same item to be retrieved + error terms of \eqref{eq:retrieval} but in  more complex space that contains all cue, contexts and items stored in the memory. We finish by noting that this equation suggests that a context which maximizes $v_j$ will increase the probability of a correct retrieval result, but many different contexts could have been used. Indeed, we need merely shift the basis in equation~\eqref{eq:Vcontext} in order to obtain a very different set of representations for the items in memory, and these would have a very different set of probabilities for recall. Thus, with a shift to a geometric space we see a way in which information might be stored and retrieved from a system based upon traditional storage mechanisms than is currently the case, all through the use of a sophisticated notion of context.

\section{Conclusion}

A strength of this approach lies in the density of information that it is likely to be able to  store. The choice of a structural approach to information storage, with a potentially infinite set of contexts, allows for memory to move from a density driven approach, where the quantity of information stored is inversely proportional to the size of the components used to store it, towards one where storage capacity is limited only by how many sensible contexts can be used to retrieve the required information. Even with a very small storage space, a wide range of representations can still be obtained from a conceptual space that takes the underlying subsymbolic structure and complexifies it according to the context in which it is accessed. 
Such an ``actualization of potential'' \cite{gabora.saab:creative} provides both creative ability and extra storage capability. Indeed, with this approach, we can start to see how the lowest, or neural level of cognition can be made redundant despite its strong dependence upon a specific structure.

While we have emphasized the background of this quantum inspired model in the field of Information Retrieval, a related line of work \cite{aerts.gabora:theory}--\cite{kitto.ramm.ea:quantum,bruza.kitto.ea:non-decomposability,aerts.broekaert.ea:case} makes use of a quantum approach to model concepts and their combinations. Thus, a growing body of literature points to the utility of the quantum formalism in modeling Information in context from both the cognitive and the computational sides of Information storage and retrieval.
This approach has also been utilized in a  preliminary approach illustrating how context might be incorporated into vector spaces described with reference to a point of view \cite{aerts.kitto.ea:similarity}. A solution which that paper shows might circumvent the apparent incompatibility between metricity and the similarity judgments that humans actually make \cite{tversky.gati:similarity}.
 
While the proposed approach is in its very early days, we feel that its incorporation of a wide range of both cognitive and computational insights makes it a highly interesting avenue to pursue as we search for new paradigms of computational memory and information storage. Future work will investigate the manner in which different contexts might interfere with specified cues to produce different probabilities of recall and hence different items from memory. It will also seek to further clarify the role of the projection operators in specifying a context space, and to extend the formalism proposed at the end of the previous section. Finally, we intend to take seriously the notion of creativity as it arises in human memory, and to investigate the manner in which a similar notion might arise in a system such as this. Such a result would bring us one step closer towards a system capable of exhibiting a true form of computational intelligence.


\section*{Acknowledgments}

This project was supported in part by the Australian Research Council Discovery grant DP1094974, the Social Sciences and Humanities Research Council of Canada, and the Fund for Scientific Research of Flanders, Belgium. Welcome support was also provided by the Marie Curie International Research Staff Exchange Scheme: Project 247590, ``QONTEXT - Quantum Contextual Information Access and Retrieval"). 


\bibliographystyle{IEEEtran}
%\bibliography{bib}



\end{document}